\documentstyle[prd,aps,epsfig,subfigure]{revtex}   
\newcommand{\be}{\begin{equation}} 
\newcommand{\ee}{\end{equation}}
\newcommand{\ba}{\begin{eqnarray}} 
\newcommand{\ea}{\end{eqnarray}}
\newcommand{\no}{\nonumber \\} 
\newcommand{\bb}{\bibitem}

\newcommand{\nb}{\mbox{$N_{B}$}}
\newcommand{\ndb}{\mbox{$n_{B}$}}

\newcommand{\s}{\mbox{$\rm{sec}$}}
\newcommand{\au}{\mbox{$\rm{AU}$}}
\newcommand{\cm}{\mbox{$\rm{cm}$}}
\newcommand{\km}{\mbox{$\rm{km}$}}
\newcommand{\fm}{\mbox{$\rm{fm}$}}
\newcommand{\m}{\mbox{$\rm{m}$}} 
\newcommand{\g}{\mbox{$\rm{g}$}}
\newcommand{\mev}{\mbox{$\rm{MeV}$}}
\newcommand{\mpc}{\mbox{$\rm{Mpc}$}}

\newcommand{\delh}{\mbox{$\delta_{H}$}}
\newcommand{\delbar}{\mbox{$\bar{\delta}$}}
\newcommand{\delbarh}{\mbox{$\bar{\delta}_{H}$}}

\newcommand{\mh}{\mbox{$M_{H}$}}

\newcommand{\bgl}{\mbox{$\biggr($}}
\newcommand{\bgr}{\mbox{$\biggr)$}}
\newcommand{\mstarh}{\mbox{$M^{\ast}_{H}$}}
\newcommand{\mch}{\mbox{$M^{c}_{H}$}}
\newcommand{\mqh}{\mbox{$M^{Q}_{H}$}}
\newcommand{\mc}{\mbox{$M_{c}$}}
\newcommand{\oqn}{\mbox{$\Omega_{QN}$}}
\newcommand{\lqn}{\mbox{$l_{QN}$}}

\newcommand{\ncb}{\mbox{$N^{c}_{B}$}}
\newcommand{\nbh}{\mbox{$N_{B-H}$}}

\newcommand{\mqcd}{\mbox{$M_{Q}$}}

\newcommand{\mqn}{\mbox{$M_{QN}$}}

\newcommand{\delbari}{\mbox{$\bar{\delta_{i}}$}}

\newcommand{\simh}{\mbox{$\sigma_{H}$}}

\newcommand{\kqh}{\mbox{$k_{H}$}}

\newcommand{\delch}{\mbox{$\delta^{c}_{H}$}}

\newcommand{\mnr}{\mbox{Mon. Not. R. Astron. Soc.}}

\newcommand{\lsim}{\lesssim}
\newcommand{\gsim}{\gtrsim}
\begin{document} 
\draft

\twocolumn[\hsize\textwidth\columnwidth\hsize\csname@twocolumnfalse\endcsname 
\title{Relics of cosmological quark-hadron phase transition} 
\author{Hee Il Kim and Bum-Hoon Lee} 
\address
{Basic Science Institute and Department of Physics, Sogang University, 
121-742, Seoul, Korea} 
\author{Chul H. Lee} 
\address 
{Department of
Physics, Hanyang University, 133-791, Seoul, Korea} 
\maketitle
\begin{abstract} 
We propose that the amplified density
fluctuations by the vanishing sound velocity effect during the
cosmological quark-hadron phase transition lead to 
quark-gluon plasma lumps decoupled from the expansion of the
universe, which may evolve to quark
nuggets (QNs). Assuming power-law spectrum of density fluctuations, we
investigate the parameter ranges for the QNs to
 play the
role of baryonic dark matter and give inhomogeneities which could 
affect big-bang
nucleosynthesis within the observational bounds of
CMBR anisotropy. The QNs can give
the strongest constraint ever found on the spectral index.
\end{abstract}
\narrowtext

]

As the temperature of the universe cools down to the critical temperature
$T_{Q} \sim 100-200 \mev$, so called quark-hadron phase transition
occurs in such a way that the high temperature quark gluon plasma (QGP)
phase transforms to the low
temperature hadronic phase. If the
phase transition is first order, it is described by the nucleation of
hadron bubbles and their growth. As hadron bubbles occupy more
than half of the whole space, the picture becomes that of QGP bubbles
shrinking in the sea of the hadronic bubbles. Witten pointed out that
the shrinking QGP bubbles in the
transition may evolve to quark nuggets (QNs) which can play the role of
baryonic dark matter \cite{wit}. Baryon concentration in the shrinking QGP
bubbles also give inhomogeneities which could affect big-bang
nucleosynthesis (BBN) \cite{bbn}. Even though recent lattice
calculations show that the
quark-hadron phase transition is weakly first-order \cite{lat},  
the mean bubble
separation $l_{b}\sim 1\cm$ based on the lattice calculations
\cite{col}
 is too small for the transition to achieve the effects mentioned above.

Recently, it was shown that the growth of the subhorizon scale
fluctuations in the quark-hadron transition is amplified because
the sound velocity during the transition vanishes
\cite{sch,jed}. Amplified subhorizon scale fluctuations give
gravitational potential wells which will trap cold dark matters (CDMs)
like axions
and primordial black holes (PBHs) \cite{sch}. Resulting CDM lumps have
size of about 
$1-10 \au$ today. Also the vanishing sound velocity effect enables,
though it needs
fine-tuned initial conditions \cite{wid}, the formation of PBHs of $1
M_{\odot}\sim 10^{33}\g$ which can
be related to MACHOs \cite{jed}. The amplified fluctuations may
produce nonlinear
lumps which are decoupled from the expansion of the background
universe. If the decoupled lumps were hadronic, however, they would be washed
out by neutrino damping \cite{neu} before the big-bang nucleosynthesis era
\cite{sch,sch98}, so
could not give any significant effect on dark matter problem as well as
the inhomogeneous big-bang nucleosynthesis (IBBN) model. In this
letter, we propose that the fluctuations amplified by the vanishing
sound velocity effect may produce nonlinear QGP lumps which will
survive till today in the form of QNs. With a
 power-law assumption on the density
fluctuation spectrum, we find possible parameter ranges for which
the QNs form to play the role of dark matter and to produce 
baryon number inhomogeneities for the IBBN within the observational
bounds of CMBR anisotropy.

The sound velocity vanishes during the phase transition as follows.  
The QGP and the hadronic phase are 
in phase equilibrium with
constant temperature $T_{Q}$ shortly after reheating. 
In the cosmological situation, chemical
potential of each fluid is negligible. Thus the pressure and the
energy density depend only on
the temperature of the system. Consider the system with a scale
smaller than the horizon size at the transition $d^{Q}_{H} \sim 10 \km$
but larger
than the mean bubble separation $l_{b}$, where the details of
bubble dynamics are not important. Before and after the transition,
the system is filled with QGP and hadronic radiation fluid,
respectively, so the sound velocity $v_{s}=\sqrt{1/3}$. As the phase
transition proceeds, while the
pressure remains constant because the two phases are in pressure
equilibrium, the energy density of the system gradually decreases. 
Therefore, the
sound velocity $v_{s}=(\partial p/\partial \rho)^{1/2}_{s}=0$ during
the phase transition \cite{sch,jed}. Since the sound
velocity vanishes, the pre-existing density fluctuations
 grow without any pressure
gradient and restoring force and the amplitudes of the fluctuations are
amplified. Superhorizon scale fluctuations are not affected by the
sound velocity transition. 

The subhorizon fluctuations which entered the horizon long before the
quark-hadron transition were damped by the neutrino
diffusion. Only the fluctuations with the comoving wavenumber $k\gsim
10^{4}\kqh$ can retain their horizon crossing amplitude $\delh$
\cite{sch98},
where $\kqh$ is the comoving wavenumber of the fluctuation
which enters the horizon at the transition. During the whole time of
the transition, the sound
velocity transition amplify the amplitudes of the subhorizon
fluctuations with $\delh$ and $k$ upto
\be
\delta^{amp}_{H}=\biggr(\frac{k}{\kqh}\biggr)^{\zeta} \delta_{H}
~~{\rm{for}}~~ \kqh\lsim k\lsim 10^{4}\kqh
\ee
where $\zeta=1$ for the bag model and
$\zeta=3/4$ for lattice calculation \cite{sch}. 
The exponent $\zeta$ is obtained for $k\gg \kqh$ but we simply
regard it valid for $k\gsim \kqh$. The fluctuations with 
$\delta^{amp}_{H}\gsim 1$, are nonlinearized by the sound velocity 
transition and are
decoupled from the expansion of the background universe. So it needs
$\delh \gsim \delch=(\kqh/k)^{\zeta}$ for the nonlinearization. Since this
condition is far weaker than $\delch \sim 1$ in ordinary case without the
sound velocity transition, the nonlinear objects with mass below the
horizon mass at the transition $\mqh \approx 10^{33}\g$
 (even PBHs with $10^{33}\g$ \cite{jed}) are formed much
more 
copiously during the quark hadron phase transition. 

What will happen as the density fluctuations grow and become
nonlinear? We propose that some nonlinearized fluctuations lead to
the formation of QNs.
The nonlinearization of fluctuations with scale $\lambda$ in the 
linear analysis can be interpreted in another sense
as the overdense regions of size $\lambda$
evolving into gravitationally bound lumps decoupled from the expansion 
of the universe. The temperature of the decoupled lumps will not
decrease any more by the expansion of the universe. Instead, the increased
pressure and energy density of the lumps by the nonlinearization (or
the gravitational binding) will
induce the transition to the QGP phase in the hadronic phase regions.
 If the overdense regions were composed of a single phase, 
the temperature of the overdense
region, in  general, would increase as the energy density
increases. For
a region with mixed phases of QGP and hadronic phases, however, the
temperature of the region does not increase until the hadronic
phase is converted entirely to the QGP phase.
The phase of the nonlinearized lump is determined by comparing two
important but uncertain time scales. One
is $t_{nonl}$, the time needed for the overdense region to become
nonlinear.
 The other is $t_{back}$, the time for the mixed phase to
return to QGP phase. What is needed for the decoupled lumps to return
to the QGP
phase is $t_{nonl}\sim t_{back}$. If $t_{back}>t_{nonl}$, the decoupled lumps
are made of almost hadronic phase so that they will be washed out by
the neutrino damping. Also, if $t_{back}<t_{nonl}$,
the lump is still in linear
regime when it is entirely converted back to the QGP phase and it
experiences
 ordinary phase transition to the hadronic phase due
to the cooling by the expansion of the universe before it becomes nonlinear.
 Therefore, it needs $t_{nonl}\sim
t_{back}$ for nonlinear QGP lump formation. This can be achieved for
the overdense regions with
$\delh \sim \delta^{c}_{H}$. For such regions $t_{nonl}\sim
\Delta t_{Q}$
 that is, roughly the time needed
for the phase transition between two phases. 
So we can conceive that the overdense regions with
$\delh \sim \delch$ evolve to nonlinear QGP lumps. Though the
overdense regions with $\delh \gg \delch$ which have $t_{nonl}\ll
\Delta t_{Q}$ may also satisfy the condition $t_{nonl}\sim t_{back}$,
 the number of such
regions is negligible as can be seen in the probability distribution
of overdense regions.

The mass and size of the QGP lump are specified by the baryon number
in the lump $N_{B}\propto k^{-3}$ as
$
\mqcd=(\frac{\nb}{\nbh})\mqh
$
where $\nbh=10^{49}(T_{Q}/100\mev)^{-3}$ is the total baryon number
contained in the horizon at the transition. For $k\approx
10^{4}\kqh$, the minimum scale relevant to the sound velocity
transition, $\nb \approx 10^{37}$.
 The QGP lump can evolve to a QN if the
 quarks in the lump become degenerate. It occurs
if the baryon density to entropy density ratio, $\ndb/s \sim 1$. 
 For comparison, the background universe has $(\ndb/s)_{bg} \simeq
3.8\times 10^{-9} (\Omega_{B}h^{2})$ where $\Omega_{B}$ is the density
fraction of the baryonic matter at present and $h$ is the Hubble
parameter in units of $100\km~\s^{-1}~\mpc^{-1}$. The resulting QN
mass $\mqn$ is
determined by how and when the lump get enormous increase in $\ndb/s$
and transform to a QN. 
Here, we do not attempt to answer
these questions but will give some qualitative descriptions. For this we
introduce a parameter $\kappa$, the QN to the QGP lump volume ratio. 
If one takes the quark matter energy density $\simeq$
the QGP energy density at the transition, then
$
\mqn=\kappa \mqcd
$.
From a chromoelectric flux-tube model (CFTM), it was found that the
baryon penetration at the surface of the QGP lump is very low and
the baryon number density in the lump increases as the surface hadronizes
 \cite{sum}. With this view, QNs may form by
increased $\ndb$ during the surface hadronization \cite{sum}.
If there were no efficient ways to get rid of entropy and the QN
formed solely by the surface hadronization (only by the increase of
$\ndb$), then $\kappa \sim (\ndb/s)_{bg}$ and $\ndb$ should be increased to about the
baryon number density of a nucleon $\sim 0.3\fm^{-3}$. However, for the
QGP lump with $\nb \lsim 10^{34}(\ndb/s)^{-1}_{bg}\sim 10^{42}
(\Omega_{B}h^{2})^{-1}$, the lump becomes smaller than the neutrino mean
free path $\l_{\nu}\sim 10 (T/100\mev)^{-5}\cm$ before reaching
$\kappa \sim (\ndb/s)_{bg}$. Then the entropy in the lump will be reduced
rapidly by neutrinos. If this occurs faster than the
surface hadronization, then $\kappa$ can be approximated as $\kappa
\sim (\nb/10^{34})^{-1}$. In fact, the entropy loss by neutrinos
also occur at distances $\lsim \l_{\nu}$ from the surface and it needs
more detailed calculations to find the condition for QN
formation. However, we simply assume $\kappa
\simeq (\nb/10^{34})^{-1}$ as the upper limit for the QN
formation. Resulting QNs have masses $\lsim 10^{18}\g$.
In the models generating large entropy before the quark-hadron
transition \cite{wei}, the QNs can be formed easily with $\kappa \sim
1$. This situation is, however, not relevant to our work because there
can be no sound velocity transition.

Once the QNs form, they are unstable to surface evaporation
\cite{eva} and
boiling \cite{boi}. Only the QNs with the baryon number larger than
the critical baryon number $\ncb$
 can survive till
today. With small baryon penetrability in the CFTM, $\ncb$ is lowered
and two results of $\ncb= 10^{39},10^{44}$ are known
\cite{sum91,bha}. 
However, as
claimed by the authors, their works overestimated the baryon
evaporation. Especially, they assumed flavor equilibrium in their
calculations. The flavor non-equilibrium will reduce $\ncb$. So, we
take $\ncb$ as a free parameter bearing in mind the above
values. The baryon number of the QNs needed to affect the IBBN is
smaller than $\ncb$, but the
difference will not be large considering the rapid evaporation rate
below $\ncb$ \cite{sum91,bha}.

Now, we have found the conditions for the QN
formation,
$\delh \gsim \delta^{c}_{H}$ and $\nb \gsim \ncb$. 
To estimate the
number of the QNs, we just count how many overdense regions 
satisfying the conditions exist. We would like to emphasize that the
number density of the QNs depends much on the details
of density fluctuation. Though the details of the phase transition may
rule out the QN formation from the shrinking QGP bubbles, whether
the QNs from the sound velocity transition can be produced
enough is another problem. One thing to note further is that the QN
 formation proposed here occurs even without the sound velocity
transition but with much severe condition $\delh \gsim 1$. The sound
velocity transition relax the condition and enables more copious QN formation.

We assume a simple power-law spectrum of density fluctuation
 with $|\delta_{k}|^{2} \propto k^{n}$, where $\delta_{k}$ is
the Fourier transform of $\delta({\bf{x}}) \equiv
(\rho({\bf{x}})-\rho_{b})/\rho_{b}$ and the spectral index $n$ is a constant. 
 The initial power spectrum of the fluctuation
amplitude is defined by the rms amplitude for a given logarithmic
interval in $k$, $\delbar^{2}(k)\equiv
k^{3}|\delta_{k}|^{2}/2\pi^{2}$. From the linear analyses, the rms
 amplitude $\delbari(k)$ at $t_{i}$, the time when the fluctuations develop,
 is related to the rms horizon crossing amplitude $\delbarh(k)$ as
follows
\ba
\delbarh(k)=\biggr(\frac{k}{k_{Hi}}\biggr)^{-2}\delbari(k)
        =\biggr(\frac{k}{k_{0}}\biggr)^{(n-1)/2}\delbarh(k_{0})
\ea
where $k_{Hi}$ is the wave number of the horizon scale at $t_{i}$
 and the subscript `0' denotes the values at
present. From the COBE measurement, $\delbarh(k_{0}) \approx 10 ^{-5}$
 and $n=1.2 \pm 0.3$ \cite{gor}. The number density of overdense regions with
$\delh \gsim \delch$ can be found by the
 Press-Schechter method \cite{pre} which is widely used to estimate the number
 density of structures like galaxies. The difference here is in the
 scale dependence of the critical amplitude $\delch \propto
 k^{-\zeta}$. The initial mass spectrum of
 the QGP lump in the range $(\mh,\mh+d\mh)$ where $\mh \propto M^{2/3}_{Q}$ is the horizon
 mass of scale with $k$ 
\ba
n(\mh)d\mh &&\propto \mh^{(\alpha-10)/4} \exp \biggr[ -\bgl \frac{\delch}
{\sqrt{2}\simh}  \bgr^{2}\biggr] d\mh \no
 &&\propto \mh^{(\alpha-10)/4} \exp \biggr[-\bgl
\frac{\mh}{\mstarh}\bgr^{\alpha/2}\biggr] d\mh
\ea
where $\alpha=n+2\zeta-1$ and
$\simh=\simh_{0}(\mh/\mh_{0})^{(1-n)/4}$ is the filtered
 amplitude of $\delbar_{H}(k)$ and $\simh_{0}=9\times 10^{-5}$ \cite{gre}.
The cutoff scale $\mstarh \ll \mch$. Due to the scale dependence of
 $\delch$, the mass spectrum has exponential cutoff
 even with $n=1$
Harrison-Zel'dovich spectrum. For comparison, $n(\mh)
\propto \mh^{(n-11)/4} \exp (-\mh^{(n-1)/2}) $ without the sound velocity
transition \cite{mp2}, so if $n=1$, the mass spectrum could have very
 broad mass
ranges with $n(\mh) \propto \mh^{-5/2}$. With the exponential cutoff,
the QN with only $\nb \simeq \ncb$ form significantly. 
So we can assume $\delta$-function type mass
spectrum and approximate the initial number fraction of
the QNs with $\ncb$ as
\be
\beta_{i}(M^{c}_{H})=\simh(M^{c}_{H})
\exp \biggr[ -\bgl \frac{\delch}{\sqrt{2}\simh(M^{c}_{H})} 
 \bgr^{2}\biggr]
\ee
which is used for calculating the number fraction of PBHs formed by
 the density fluctuations with blue-shifted spectrum. 

Since the QNs can be regarded as pressureless dust, the
density fraction at present is
\be
\Omega_{QN}(t_{0})=\Omega_{QN}(t_{eq})=
\frac{\rho_{i}}{\rho_{eq}}
\biggr( \frac{M^{c}_{QN}}{\mc}\biggr) \biggr(\frac{a(t_{eq})}{a(t_{i})}
\biggr)
^{-3}\beta_{i}
\ee
where the subscript `$eq$' represents the values at matter-radiation
equal time. $M_{c}$ is the mass contained in the overdense region with
$k_{c}$ at $t_{i}$.
The density fraction can be arranged into
\be
\Omega_{QN}(t_{0})=\kappa \biggr(\frac{T_{Q}}{T_{eq}}\biggr) \beta_{i}~~.
\ee
Fig. 1 shows the relation between $\kappa$ and $\ncb$ to satisfy
$\oqn=1$ for possible
range of the spectral index, $0.9\leq n \leq 1.5$. It seems impossible
for QNs to be formed solely by the surface hadronization
($\kappa \simeq (\ndb/s)_{bg}$). The QGP lumps become smaller than $l_{\nu}$
when $\kappa=(\ncb/10^{34})^{-1}$ (the bold line in Fig. 1). So, in
the left side of the bold line, the QGP lumps can evolve to QNs
($\ndb/s \sim 1$) by further losing their entropy by neutrinos. The
QN formation is relevant for only $\ncb \lsim 10^{42}$. For
$n\approx 1$, it needs $\ncb \lsim 10^{38}$ for $\zeta=1
$, which is smaller than the CFTM results \cite{sum91,bha}. With $\zeta=0.75$, 
the QN formation is possible if $n\gsim 1.1$. Assuming
 $\kappa=(\ncb/10^{34})^{-1}$, the upper limits on $n$ are found from
the condition $\oqn \leq 1$. If $\ncb \lsim 10^{40}$, the QN can
give strongest constraints ever found, $0.9\lsim n_{upper}\lsim 1.2$
 (the bold lines in Fig. 2).  PBHs
can give at best $n_{upper}\gsim 1.23$ even weaker as the reheating
temperature increases \cite{gre,mp3}. 
Note that the upper limits on $n$ do not depend on $t_{i}$ or the
reheating temperature in inflationary models. Demanding less $\oqn$,
the parameter ranges become broader and the constraints on $n$
becomes stronger.

Since QNs have positive electric surface potential of order 
$\sim \mev$, they absorb only neutrons. This reduces the neutron to
proton ratio, so lowering the helium
production. Not to violate the observations, the QNs should have size
$\gsim 10^{-6} \cm$ assuming $\oqn =1$
and perfect baryon penetrability (even smaller with smaller
penetrability) \cite{mad}. The corresponding $\ncb \simeq 10^{13}$, so
the QNs considered here easily satisfy the condition. 
More significant effects on the IBBN can be induced by the evaporation
of the QNs. The minimal
requisition for the IBBN is that the mean separation between QNs 
should be larger than proton diffusion length when the BBN starts. It
requires the mean separation of QNs at the transition, $\lqn
\gsim 3\m$ \cite{suh} ruling out the QNs from the
shrinking QGP bubbles ($\l_{b} \approx 1\cm$). We find
\be
\lqn \approx n^{-1/3}_{QN}
\approx 10^{8} \bgl\frac{N^{c}_{B}}{N_{B-H}}\bgr^{1/3}
\beta^{-1/3}_{i} \m~~.
\ee
The upper values of $n$ for the IBBN ($\l_{QN} \approx 3 \m$) are shown
in Fig. 2. It can be seen that the the minimal condition is easily
satisfied. The closure condition
$\oqn \leq 1$ (the bold lines in Fig. 2) corresponds to demanding 
$\lqn \gsim 50\m$. It goes to $\lqn \gsim 100\m$ demanding 
$\oqn \leq 0.1$ not shown
in Fig. 2. Interestingly, demanding $0.1\lsim \oqn \lsim 1$, $\lqn$ lies in the
ranges for the IBBN to be effective \cite{suh,jed94}. So, the QNs 
can contribute the density of the universe in the
form of baryonic dark matter as well as the IBBN. The enhanced
heavy element formation can be the signature for the QNs \cite{jed94}.

In summary, we propose that QNs can be formed by the sound velocity
transition during the quark-hadron phase transition. The difficulties
of the QNs formed by the shrinking QGP bubbles can be
avoided by the QNs from the sound velocity transition because their
formation depends much on the details of density fluctuations. So,
they can dominate the density of the universe
$\oqn \approx 1$ and affect the standard BBN. Also, the QN
formation can strongly constrain the spectral index. Our
analyses are so far rather qualitative and include undetermined parameters
such as $\kappa$ and $\ncb$. It needs further systematic analyses to get more
quantitative results. After all, the sound velocity transition effect
can make the QN to play an important role during the evolution of the
universe.


This work was supported in part by the Korean Ministry of Education
and the KOSEF. HIK was also supported by the KOSEF.

\begin{figure}[htbp]
\epsfig{file=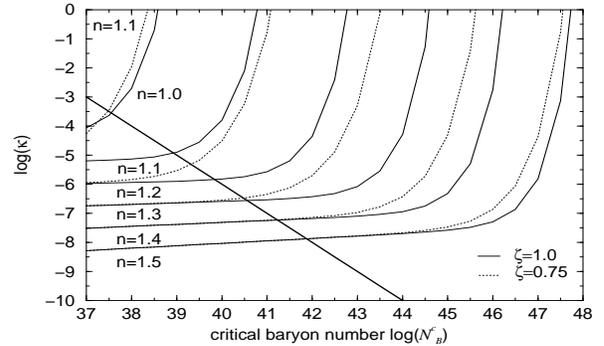,width=0.9\linewidth,height=0.2\textheight}
\caption{The parameter ranges for $\kappa$ and $\ncb$ within the
  observational bounds of $n=1.2\pm 0.3$. for $T_{Q}=150 \mev$}
\end{figure}
\begin{figure}[htbp]
\epsfig{file=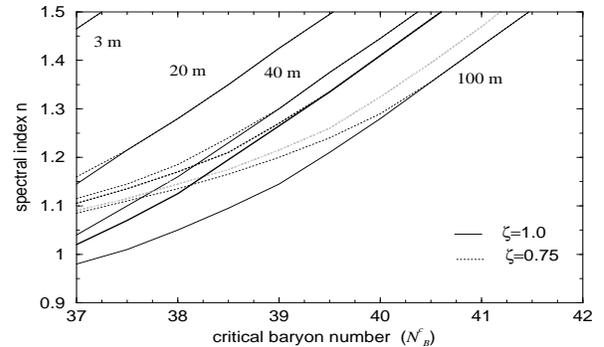,width=0.9\linewidth,height=0.2\textheight}
\caption{The relation between the spectral index and the inhomogeneity
  scale $l_{QN}$ for $T_{Q}=150\mev$. The bold lines are the upper
  limits on $n$ found from $\oqn=1$ with $\kappa=(10^{34}/\ncb)$}
\end{figure}

\end{document}